\documentclass[conference]{IEEEtran}
\IEEEoverridecommandlockouts

\usepackage{cite}
\usepackage{amsmath,amssymb,amsfonts}
\usepackage{algorithmic}
\usepackage{graphicx}
\usepackage{textcomp}
\usepackage{xcolor}
\usepackage{booktabs}
\usepackage{multirow}
\usepackage{algorithm}
\usepackage{bm}
\usepackage{subcaption}

\def\BibTeX{{\rm B\kern-.05em{\sc i\kern-.025em b}\kern-.08em
    T\kern-.1667em\lower.7ex\hbox{E}\kern-.125emX}}

\begin{document}

\title{DriftDecode: One-Step Wireless Image Decoding via Drifting-Inspired Detail Recovery}

\author{Jingwen Fu, Ming Xiao, Mikael Skoglund\\
KTH Royal Institute of Technology \\
E-mail:~\{jingwenf, mingx, skoglund\}@kth.se
}

\maketitle

\begin{abstract}
Generative receivers for wireless image transmission can improve reconstruction quality, but diffusion-based and flow-based decoding relies on iterative inference and therefore incurs substantial latency. In wireless image transmission, however, the received signal already preserves the coarse structure of the source image. Wireless decoding is therefore better viewed as a recovery task than as image generation from scratch, and the main challenge lies in restoring channel-impaired details. Motivated by this recovery-oriented perspective, this paper proposes DriftDecode, a signal-to-noise ratio (SNR)-conditioned one-step decoder for wireless image reconstruction. DriftDecode couples a one-step U-Net decoder with a drift-inspired instance-level texture loss. The loss reformulates the drifting-field mechanism from generative drifting models in perceptual feature space, guiding each reconstructed local feature toward its spatially aligned ground-truth counterpart while suppressing mismatched textures. Experiments on DIV2K and MNIST under additive white Gaussian noise (AWGN) and Rayleigh fading channels show a favorable quality-latency tradeoff. DriftDecode achieves 30~ms decoding latency, providing a 4.8$\times$ speedup over a 10-step flow-matching decoder, while consistently outperforming MSE-only training and yielding up to 1.13~dB PSNR gain on MNIST under Rayleigh fading. These results support recovery-oriented one-step decoding as an effective alternative to iterative generative decoding for low-latency wireless image transmission.
\end{abstract}


\section{Introduction}
\label{sec:intro}

Deep joint source-channel coding (DeepJSCC) has advanced wireless image transmission through end-to-end optimization, thereby mitigating the cliff effect of separated source and channel coding schemes~\cite{bourtsoulatze2019deep,xu2021wireless,yang2022ofdm}. More recently, generative receivers have been introduced to further improve reconstruction quality by exploiting learned image priors. Representative examples include Conditional Diffusion Distortion Minimization (CDDM)~\cite{wu2024cddm} and Land-then-Transport (LTT)~\cite{fu2026land}, which build on diffusion and flow-matching formulations~\cite{lipman2023flow}. Although these receivers provide strong reconstruction fidelity, they require repeated decoder evaluations and therefore incur substantial receiver-side latency, which is undesirable in delay-sensitive wireless systems~\cite{fu2025computation}. By contrast, conventional one-step DeepJSCC decoders are computationally efficient but, when trained mainly with pixel-domain losses, often produce oversmoothed reconstructions with limited perceptual detail~\cite{blau2018perception}. The resulting tradeoff between reconstruction quality and decoding latency motivates the design of one-step receivers with improved detail restoration capability.

The above tradeoff is closely related to the nature of the wireless decoding problem. Unconditional generative models must synthesize both global structure and fine texture from random noise, which naturally favors iterative refinement. Wireless image decoding, however, is fundamentally a \emph{recovery} problem: the received signal already preserves substantial structural information about the transmitted image~\cite{bourtsoulatze2019deep}, and the receiver mainly needs to compensate for channel-impaired details such as high-frequency textures and local patterns. From this perspective, the multi-step structure-building process of diffusion- and flow-based receivers is not fully aligned with the wireless reconstruction task. The key challenge is instead to design a one-step decoding objective that explicitly promotes detail restoration beyond pixel-domain mean squared error (MSE)~\cite{johnson2016perceptual}.

Motivated by this observation, this paper proposes \emph{DriftDecode}, an SNR-conditioned one-step decoder for wireless image reconstruction. DriftDecode combines a one-step U-Net receiver with a drift-inspired instance-level texture loss. Specifically, the drifting-field mechanism from drifting models~\cite{deng2026drifting} is reformulated for paired wireless image reconstruction in perceptual feature space, such that each reconstructed local feature is guided toward its spatially aligned ground-truth counterpart while mismatched spatial features are suppressed through softmax competition. The resulting loss provides position-aware supervision tailored to channel-impaired detail recovery. To the best of our knowledge, DriftDecode is the first work to adapt the drifting-field mechanism to wireless image reconstruction.

\begin{figure*}[t]
\centering
\includegraphics[width=0.85\textwidth]{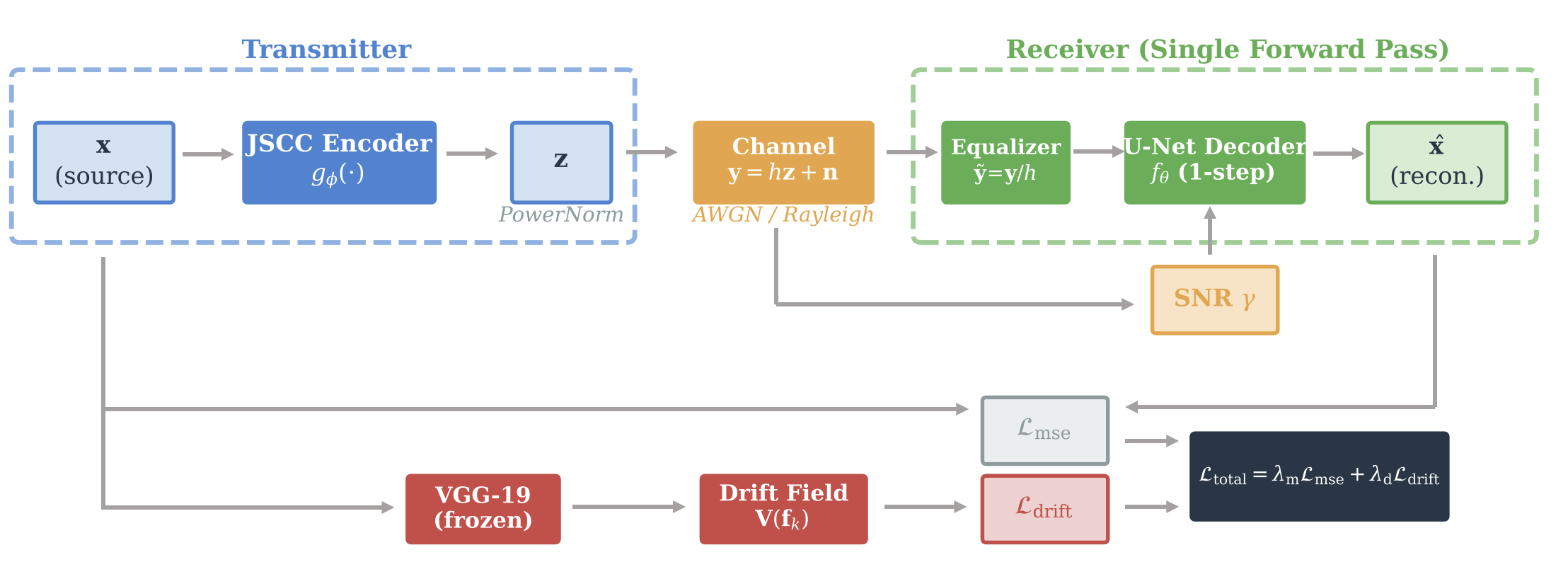}
\caption{System architecture of the proposed DriftDecode framework.}
\label{fig:system}
\end{figure*}

The main contributions of this work are summarized as follows:
\begin{enumerate}
    \item A recovery-oriented formulation of wireless image decoding is introduced, and an SNR-conditioned one-step decoder is developed accordingly. The proposed formulation emphasizes channel-impaired detail restoration rather than iterative image generation.

    \item A drift-inspired instance-level texture loss is proposed by reformulating the drifting-field mechanism for paired wireless image reconstruction. Operating in perceptual feature space, the proposed loss provides position-aware supervision for aligned local detail recovery.

    \item Experiments on DIV2K and MNIST under AWGN and Rayleigh fading channels demonstrate that DriftDecode achieves a favorable quality-latency tradeoff, consistently improving upon MSE-only training while reducing latency by $4.8\times$ relative to a 10-step flow-matching decoder.
\end{enumerate}

\section{System Model and Problem Formulation}
\label{sec:system}

As shown in Fig \ref{fig:system}, an end-to-end wireless image transmission system is considered, in which a source image $\mathbf{x} \in \mathbb{R}^{C \times H \times W}$ is transmitted over a noisy wireless channel. The system consists of a DeepJSCC encoder, the physical channel, and a one-step receiver.

A DeepJSCC encoder $g_\phi(\cdot)$ maps the source image to a sequence of complex-valued channel symbols:
\begin{equation}
    \mathbf{z} = g_\phi(\mathbf{x}),
\end{equation}
where $\mathbf{z} \in \mathbb{C}^M$ denotes the transmitted channel-symbol sequence and satisfies the average transmit-power constraint $\frac{1}{M}\mathbb{E}[\|\mathbf{z}\|^2] \leq P_z$. The bandwidth constraint is characterized by the channel bandwidth ratio (CBR), defined as $\mathrm{CBR}=M/(CHW)$, where $C$, $H$, and $W$ denote the number of image channels, image height, and image width, respectively.

The channel input-output relationship is given by
\begin{equation}
   \mathbf{y} = h\mathbf{z} + \mathbf{n},
\end{equation}
where $\mathbf{y} \in \mathbb{C}^M$ denotes the received signal and $\mathbf{n} \sim \mathcal{CN}(\mathbf{0}, \sigma_n^2 \mathbf{I})$ is additive white Gaussian noise with variance $\sigma_n^2 = P_z \cdot 10^{-\mathrm{SNR}_{\mathrm{dB}}/10}$. The scalar $h$ denotes the channel coefficient. For the AWGN channel, $h=1$, whereas for the Rayleigh fading channel, $h \sim \mathcal{CN}(0,1)$. Assuming perfect channel state information (CSI) at the receiver, zero-forcing equalization is applied to obtain $\tilde{\mathbf{y}}=\mathbf{y}/h$, which yields an effective per-sample SNR of $\gamma_{\mathrm{eff}} = |h|^2 \cdot \mathrm{SNR}$~\cite{fu2026land}. For notational simplicity, $\tilde{\mathbf{y}}$ is used to denote the equalized decoder input, with $\tilde{\mathbf{y}}=\mathbf{y}$ for the AWGN channel.

Since $\tilde{\mathbf{y}}$ retains substantial information about the source image $\mathbf{x}$, the receiver-side reconstruction network is modeled as a parameterized one-step decoder
\begin{equation}
    \hat{\mathbf{x}} = f_\theta(\tilde{\mathbf{y}}, \gamma),
\end{equation}
where $\gamma$ denotes the SNR-related conditioning variable, i.e., $\gamma=\mathrm{SNR}$ for the AWGN channel and $\gamma=\gamma_{\mathrm{eff}}$ for the Rayleigh fading channel after equalization.

The learning objective is to optimize $(\phi,\theta)$ by minimizing the expected reconstruction loss over the source distribution and channel realizations:
\begin{equation}
\min_{\phi,\theta}\ 
\mathbb{E}_{\mathbf{x},h,\mathbf{n}}
\big[
\mathcal{L}\big(\mathbf{x},\hat{\mathbf{x}}\big)
\big].
\label{eq:sys_objective}
\end{equation}
The loss design is detailed in Section~\ref{sec:method}.

\section{Proposed DriftDecode Method}
\label{sec:method}

This section first introduces a drift-inspired detail-aware loss that reformulates the original drifting-field mechanism for paired, instance-level texture recovery in Section~\ref{subsec:driftloss}. It then describes an SNR-conditioned one-step U-Net decoder for low-latency reconstruction across channel qualities in Section~\ref{subsec:arch}. The training objective and inference procedure are summarized in Section~\ref{subsec:train}.

\subsection{Detail-Aware Drifting Loss for Instance-Level Texture Recovery}
\label{subsec:driftloss}

\begin{figure*}[t]
    \centering
    \includegraphics[width=0.9\textwidth]{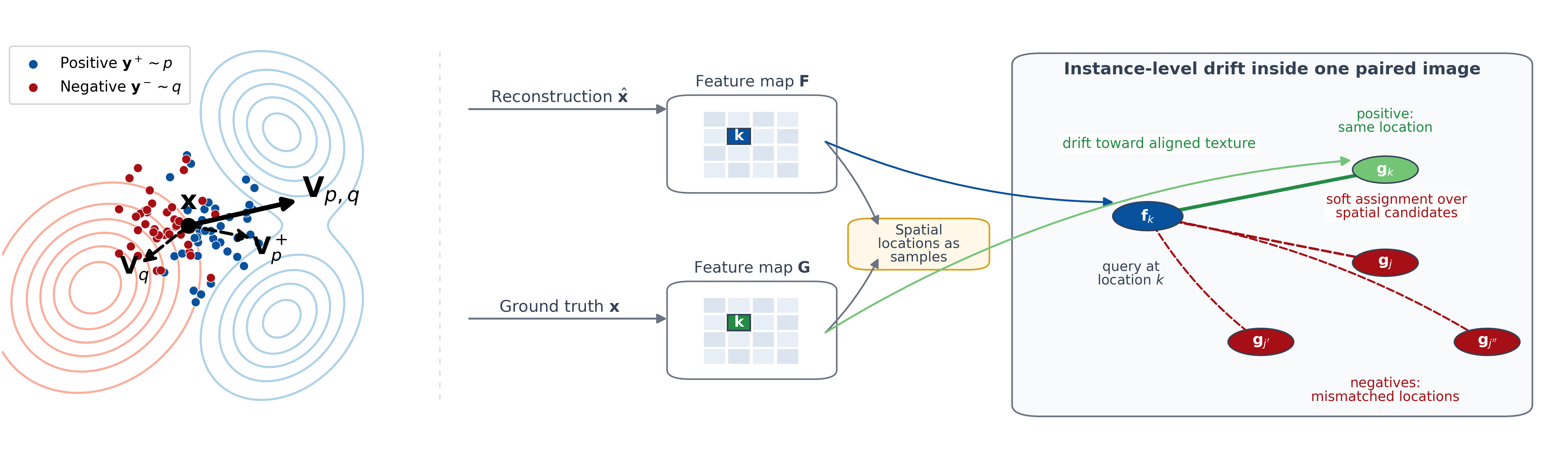}
    \caption{Comparison between the original drifting model and the proposed instance-level drift. Left: distribution-level drifting estimated from batch samples. Right: instance-level drifting constructed from spatially aligned feature locations within one image pair.}
    \label{fig:drift_compare}
\end{figure*}

\subsubsection{From generative drifting to paired recovery}
The drifting model~\cite{deng2026drifting} is a one-step generative framework that trains a generator through a fixed-point update. This update is driven by a ``drifting field'' $\mathbf{V}_{p,q_\theta}(\cdot)$, which gives the direction that moves a generated sample toward the real data distribution. The training loss is
\begin{equation}
    \mathcal{L}_{\mathrm{gen}}
    = \mathbb{E}_{\bm{\epsilon}}
    \Big[
    \big\|
    f_{\theta}(\bm{\epsilon}) -
    \text{sg}\big(f_{\theta}(\bm{\epsilon}) + \mathbf{V}_{p,q_\theta}(f_{\theta}(\bm{\epsilon}))\big)
    \big\|_2^2
    \Big],
    \label{eq:drift_gen_fp}
\end{equation}
where $\bm{\epsilon}$ denotes the initial noise input, $f_{\theta}(\bm{\epsilon})$ is the sample produced by the network, $\mathbf{V}_{p,q_\theta}(\cdot)$ is the drifting field that implements an attraction--repulsion mechanism in which samples from the target distribution $p$ attract the current prediction while samples from the model distribution $q_\theta$ repel it, and $\mathrm{sg}(\cdot)$ denotes stop-gradient.

Equation~\eqref{eq:drift_gen_fp} trains the network to match a stop-gradient drifted target, i.e., the current output after one update along $\mathbf{V}_{p,q_\theta}$. As shown in Fig.~\ref{fig:drift_compare} (left), the original drifting field is estimated from many independent samples and thus operates at the distribution level. In wireless image decoding, however, paired supervision $(\tilde{\mathbf{y}},\mathbf{x})$ is available, and the received signal usually preserves the coarse image structure. As a result, the main reconstruction errors are often local, such as missing textures and weakened high-frequency details. We therefore reformulate the drifting mechanism at the instance level, using local feature interactions within a single image pair to guide detail recovery.

\subsubsection{Instance-level drift field in perceptual feature space}

To instantiate this idea, we construct the drift field in a deep perceptual feature space rather than in the raw pixel space. This choice provides a large set of local samples through spatial feature locations and captures semantically meaningful texture variations more effectively than raw pixels~\cite{johnson2016perceptual}. 

First, as shown in Fig.~\ref{fig:drift_compare} (right), let $\phi_\ell(\cdot)$ be a frozen VGG-19 feature extractor~\cite{simonyan2015vgg}. For a selected layer $\ell\in\mathcal{L}$, we extract the feature maps of both the reconstructed image $\hat{\mathbf{x}}$ and the ground truth $\mathbf{x}$ as $\mathbf{F}^\ell$ and $\mathbf{G}^\ell$:
\begin{equation}
\mathbf{F}^\ell=\phi_\ell(\hat{\mathbf{x}})\in\mathbb{R}^{C_\ell\times H_\ell\times W_\ell},
\quad
\mathbf{G}^\ell=\phi_\ell(\mathbf{x})\in\mathbb{R}^{C_\ell\times H_\ell\times W_\ell}.
\end{equation}
By flattening the spatial dimensions into $K_\ell = H_\ell W_\ell$ locations, each spatial position $k\in\{1,\dots,K_\ell\}$ is treated as an individual sample from the local distribution. To ensure stable distance metrics, $\ell_2$-normalization is applied to these feature vectors:
\begin{equation}
    \mathbf{f}_k = \frac{\mathbf{F}^\ell_{:,k}}{\|\mathbf{F}^\ell_{:,k}\|_2+\epsilon},
    \qquad
    \mathbf{g}_k = \frac{\mathbf{G}^\ell_{:,k}}{\|\mathbf{G}^\ell_{:,k}\|_2+\epsilon}.
\end{equation}
For notational simplicity, the layer index $\ell$ is omitted below when the context is clear.

Next, for a given reconstructed query $\mathbf{f}_k$ at location $k$, its corresponding \emph{positive} target is defined as the ground-truth feature at the exact same spatial location,
\begin{equation}
    \mathbf{y}_k^{+} \triangleq \mathbf{g}_k.
\end{equation}
Conversely, the \emph{negative} targets are defined as all ground-truth features from mismatched spatial locations,
\begin{equation}
    \mathcal{Y}_k^{-} \triangleq \{\mathbf{g}_j\}_{j\in\{1,\dots,K_\ell\}\setminus\{k\}}.
\end{equation}
Combining the positive and negative targets yields the candidate set:
\begin{equation}
    \mathcal{Y}_k \triangleq \{\mathbf{y}_k^{+}\}\cup \mathcal{Y}_k^{-}.
\end{equation}
This formulation turns texture recovery into selecting the correct aligned texture while suppressing distractors from other locations. To determine the drift direction from this candidate set, the affinity between a query and each candidate is measured by a temperature-scaled negative squared distance
\begin{equation}
d_\tau(\mathbf{f}_k,\mathbf{y})=-\frac{\|\mathbf{f}_k-\mathbf{y}\|_2^2}{\tau},\qquad \mathbf{y}\in\mathcal{Y}_k,
\end{equation}
and a candidate distribution is then obtained by softmax over $\mathcal{Y}_k$:
\begin{equation}
a_\tau(k,\mathbf{y})=\frac{\exp(d_\tau(\mathbf{f}_k,\mathbf{y}))}{\sum_{\mathbf{y}'\in\mathcal{Y}_k}\exp(d_\tau(\mathbf{f}_k,\mathbf{y}'))}.
\label{eq:softmax_over_candidates}
\end{equation}
Because $\hat{\mathbf{x}}$ and $\mathbf{x}$ are spatially aligned, the positive pair $(\mathbf{f}_k,\mathbf{g}_k)$ typically has higher affinity than mismatched pairs and therefore receives larger probability mass. At the same time, all candidates share the same normalization denominator, so mismatched locations compete with the aligned target and implicitly induce repulsion when an incorrect location becomes overly similar.

Without additional normalization, the query-to-target softmax alone may assign multiple queries to a single dominant ground-truth feature, causing many-to-one matching. To prevent this, a symmetric target-to-query normalization is applied:
\begin{equation}
b_\tau(\mathbf{f}_k,\mathbf{y})=
\frac{\exp(d_\tau(\mathbf{f}_k,\mathbf{y}))}{\sum_{k'=1}^{K_\ell}\exp(d_\tau(\mathbf{f}_{k'},\mathbf{y}))}.
\label{eq:softmax_k}
\end{equation}
The joint weight $w_\tau$ is then formulated as:
\begin{equation}
\begin{aligned}
    \tilde{w}_\tau(\mathbf{f}_k,\mathbf{y})
    &= \sqrt{a_\tau(k,\mathbf{y})\,b_\tau(\mathbf{f}_k,\mathbf{y})},\\
    w_\tau(\mathbf{f}_k,\mathbf{y})
    &= \frac{\tilde{w}_\tau(\mathbf{f}_k,\mathbf{y})}
    {\sum_{\mathbf{y}'\in\mathcal{Y}_k}\tilde{w}_\tau(\mathbf{f}_k,\mathbf{y}')}.
\end{aligned}
    \label{eq:sym_weight}
\end{equation}
Here, $a_\tau(\cdot,\cdot)$ measures query-to-target proximity, while $b_\tau(\cdot,\cdot)$ discourages many-to-one matching by downweighting targets claimed by closer queries. Taking the geometric mean balances these two factors so that a target receives a large weight only when it is favored from both the query-to-target and target-to-query perspectives. The normalized weight $w_\tau$ therefore defines a normalized soft assignment over $\mathcal{Y}_k$. The temperature-specific drift direction is the residual between $\mathbf{f}_k$ and that $\tau$-dependent centroid:
\begin{equation}
    \mathbf{V}_\tau(\mathbf{f}_k) = \sum_{\mathbf{y}\in\mathcal{Y}_k} w_\tau(\mathbf{f}_k,\mathbf{y})\,\mathbf{y} - \mathbf{f}_k.
    \label{eq:inst_drift_tau}
\end{equation}

Finally, since image textures exist across multiple granularities, a single temperature $\tau$ is insufficient. The drift field is computed across a set of temperatures $\mathcal{T}$ and aggregated with per-temperature normalization:
\begin{equation}
    \mathbf{V}(\mathbf{f}_k)=
    \sum_{\tau\in\mathcal{T}}
    \frac{\mathbf{V}_\tau(\mathbf{f}_k)}{\sqrt{\frac{1}{K_\ell}\sum_{k'=1}^{K_\ell}\|\mathbf{V}_\tau(\mathbf{f}_{k'})\|_2^2+\epsilon}}.
    \label{eq:inst_drift_agg}
\end{equation}
This normalization ensures that drift forces from different texture scales contribute equally to the final update direction.

\subsubsection{Drift loss}
Following the fixed-point training objective of drifting models in Eq.~\eqref{eq:drift_gen_fp}, and using the proposed drift field $\mathbf{V}(\cdot)$ defined in Eq.~\eqref{eq:inst_drift_agg}, the one-step training objective regresses $\mathbf{f}_k$ to its drifted target $\mathbf{f}_k+\mathbf{V}(\mathbf{f}_k)$ while stopping gradients through the target:
\begin{equation}
    \mathcal{L}_{\mathrm{drift}}
    =
    \frac{1}{|\mathcal{L}|}\sum_{\ell\in\mathcal{L}}
    \frac{1}{K_\ell}\sum_{k=1}^{K_\ell}
    \left\|
    \mathbf{f}_k -
    \text{sg}\big(\mathbf{f}_k+\mathbf{V}(\mathbf{f}_k)\big)
    \right\|_2^2.
    \label{eq:dd_drift_loss}
\end{equation}
For a batch of training data, Eq.~\eqref{eq:dd_drift_loss} is computed separately for each image using only its own spatial locations, and the resulting per-image losses are then averaged across the batch. In contrast to standard perceptual losses, $\mathcal{L}_{\mathrm{drift}}$ provides an explicit, position-aware update direction for spatially aligned texture recovery: each query is pulled toward its corresponding ground-truth feature and pushed away from mismatched locations.

\subsection{SNR-Conditioned One-Step Decoder}
\label{subsec:arch}

The drift loss in Section~\ref{subsec:driftloss} provides a position-aware training signal for detail recovery. The one-step decoder $f_\theta(\cdot)$ that realizes this objective is instantiated as a multi-scale U-Net~\cite{ronneberger2015unet} with residual blocks and skip connections. The equalized complex symbols are converted to a real-valued tensor before entering the network, with real and imaginary parts stacked as separate channels. To better exploit channel CSI, the decoder is conditioned on the SNR via FiLM~\cite{perez2018film}, and the SNR embedding is injected into every residual block.

Specifically, the scalar SNR $\gamma$ is embedded using sinusoidal features and a multi-layer perceptron (MLP) to obtain a conditioning vector $\mathbf{e}_\gamma$:
\begin{equation}
    \mathbf{e}_\gamma=\mathrm{MLP}\Big(\big[\cos(\omega_j\gamma),\ \sin(\omega_j\gamma)\big]_{j=1}^{D}\Big).
    \label{eq:snr_embed}
\end{equation}
Given an intermediate feature map $\mathbf{h}$, FiLM applies channel-wise affine modulation after normalization:
\begin{equation}
    \mathbf{h}' = \bm{\alpha}(\mathbf{e}_\gamma)\odot \mathrm{Norm}(\mathbf{h}) + \bm{\beta}(\mathbf{e}_\gamma),
    \label{eq:film}
\end{equation}
where $\bm{\alpha}(\cdot)$ and $\bm{\beta}(\cdot)$ are learned projections and are broadcast over spatial dimensions. This yields a continuous family of decoding behaviors over SNR, allowing a single model to adapt detail recovery strength across the full operating range.

\subsection{Training Objective and Inference}
\label{subsec:train}

DriftDecode is trained by combining a pixel-domain reconstruction term with the proposed drift regularization:
\begin{equation}
    \mathcal{L} =
    \lambda_{\mathrm{mse}}\|\mathbf{x}-\hat{\mathbf{x}}\|_2^2 +
    \lambda_{\mathrm{drift}}\,\mathcal{L}_{\mathrm{drift}},
    \label{eq:dd_total_loss}
\end{equation}
where $\lambda_{\mathrm{mse}}$ and $\lambda_{\mathrm{drift}}$ are weights for the MSE loss term and the drift loss term. The training procedure is summarized in Algorithm~\ref{alg:training}. At inference time, DriftDecode performs a single forward pass of the neural decoder $f_\theta(\tilde{\mathbf{y}},\gamma)$ without iterative refinement, enabling one-step low-latency deployment while explicitly targeting texture restoration. 

\begin{algorithm}[!h]
\caption{DriftDecode Training Process}
\label{alg:training}
\begin{algorithmic}[1]
\REQUIRE Dataset $\mathcal{D}$, JSCC encoder $g_\phi$ (trainable), decoder $f_\theta$ (trainable), VGG feature extractor $\phi$ (frozen), temperatures $\mathcal{T}$, layers $\mathcal{L}$, weights $\lambda_{\mathrm{mse}},\lambda_{\mathrm{drift}}$
\FOR{each mini-batch $\{\mathbf{x}_i\}_{i=1}^B$}
    \STATE Sample $\gamma_i \sim \mathcal{U}[0,20]$~dB, encode $\mathbf{z}_i \leftarrow g_\phi(\mathbf{x}_i)$, transmit through the channel, and equalize to obtain $\tilde{\mathbf{y}}_i$
    \STATE Decode in one step: $\hat{\mathbf{x}}_i \leftarrow f_\theta(\tilde{\mathbf{y}}_i,\gamma_i)$
    \STATE Compute $\mathcal{L}_{\mathrm{mse}} \leftarrow \frac{1}{B}\sum_i \|\mathbf{x}_i-\hat{\mathbf{x}}_i\|_2^2$
    \STATE For each image $i$ and layer $\ell$, extract $\mathbf{F}_i^\ell,\mathbf{G}_i^\ell$, flatten and normalize the spatial features, and compute $\mathbf{V}(\mathbf{f}_{i,k})$ via Eqs.~\eqref{eq:softmax_over_candidates}--\eqref{eq:inst_drift_agg}
    \STATE Average the resulting per-image drift loss in Eq.~\eqref{eq:dd_drift_loss} over the mini-batch to obtain $\mathcal{L}_{\mathrm{drift}}$
    \STATE Update $(\phi,\theta)$ using $\mathcal{L} \leftarrow \lambda_{\mathrm{mse}}\mathcal{L}_{\mathrm{mse}}+\lambda_{\mathrm{drift}}\mathcal{L}_{\mathrm{drift}}$
\ENDFOR
\end{algorithmic}
\end{algorithm}

\section{Numerical Results}
\label{sec:experiments}

\begin{figure*}[t]
    \centering
    \begin{subfigure}[t]{0.24\textwidth}
        \centering
        \includegraphics[width=\textwidth]{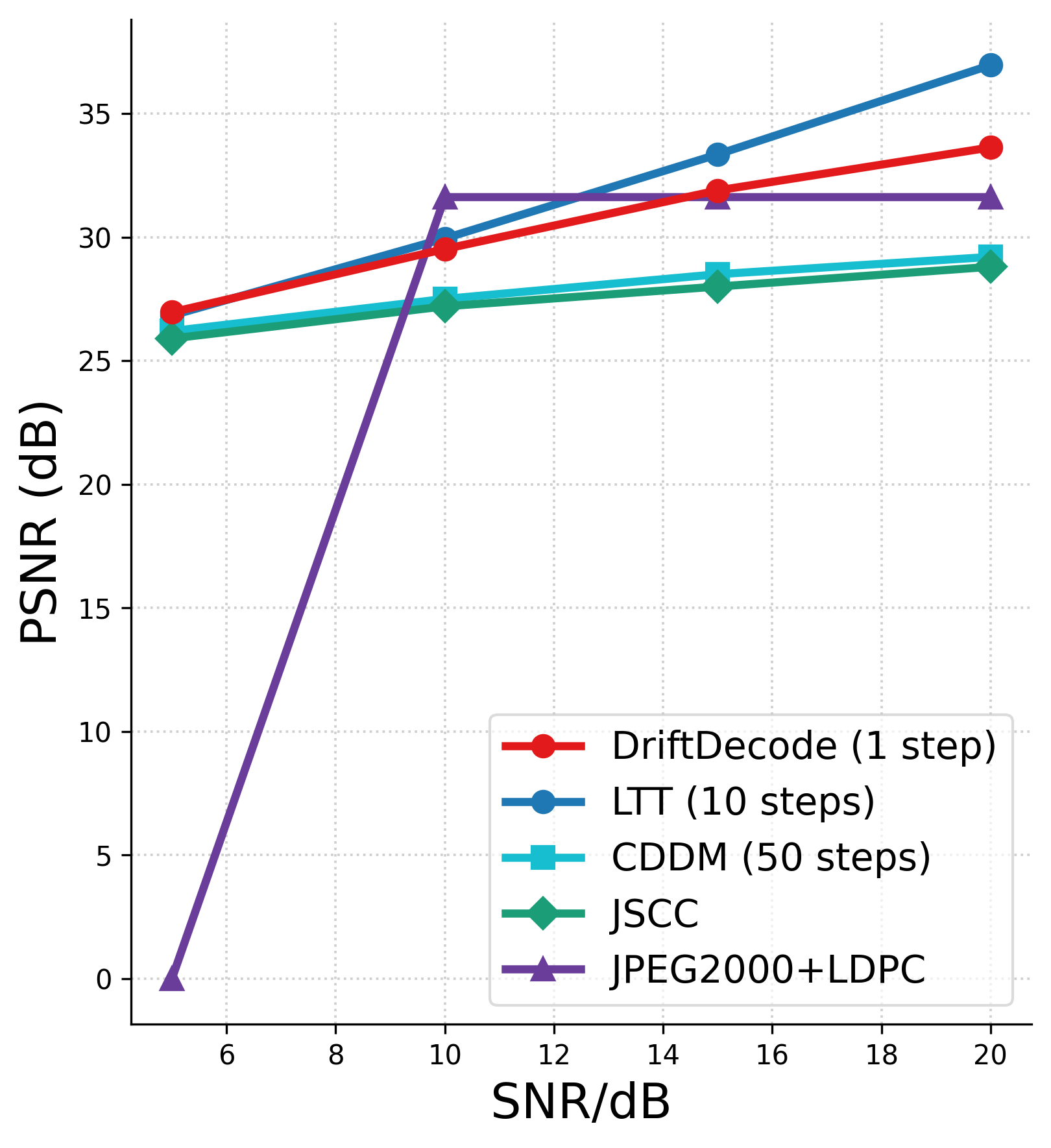}
        \caption{AWGN - PSNR}
        \label{fig:baseline_awgn_psnr}
    \end{subfigure}
    \hfill
    \begin{subfigure}[t]{0.24\textwidth}
        \centering
        \includegraphics[width=\textwidth]{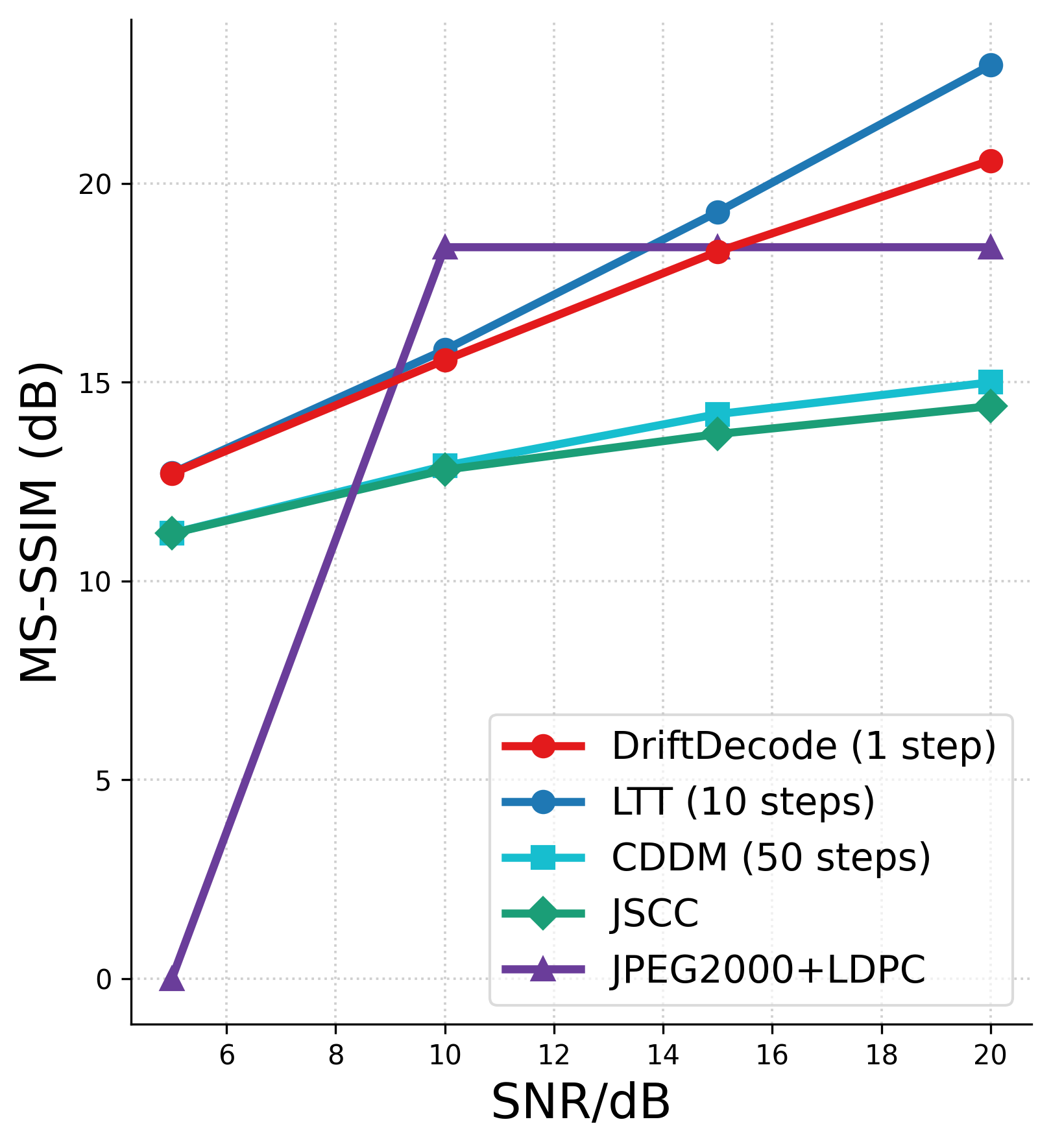}
        \caption{AWGN - MS-SSIM}
        \label{fig:baseline_awgn_msssim}
    \end{subfigure}
    \hfill
    \begin{subfigure}[t]{0.24\textwidth}
        \centering
        \includegraphics[width=\textwidth]{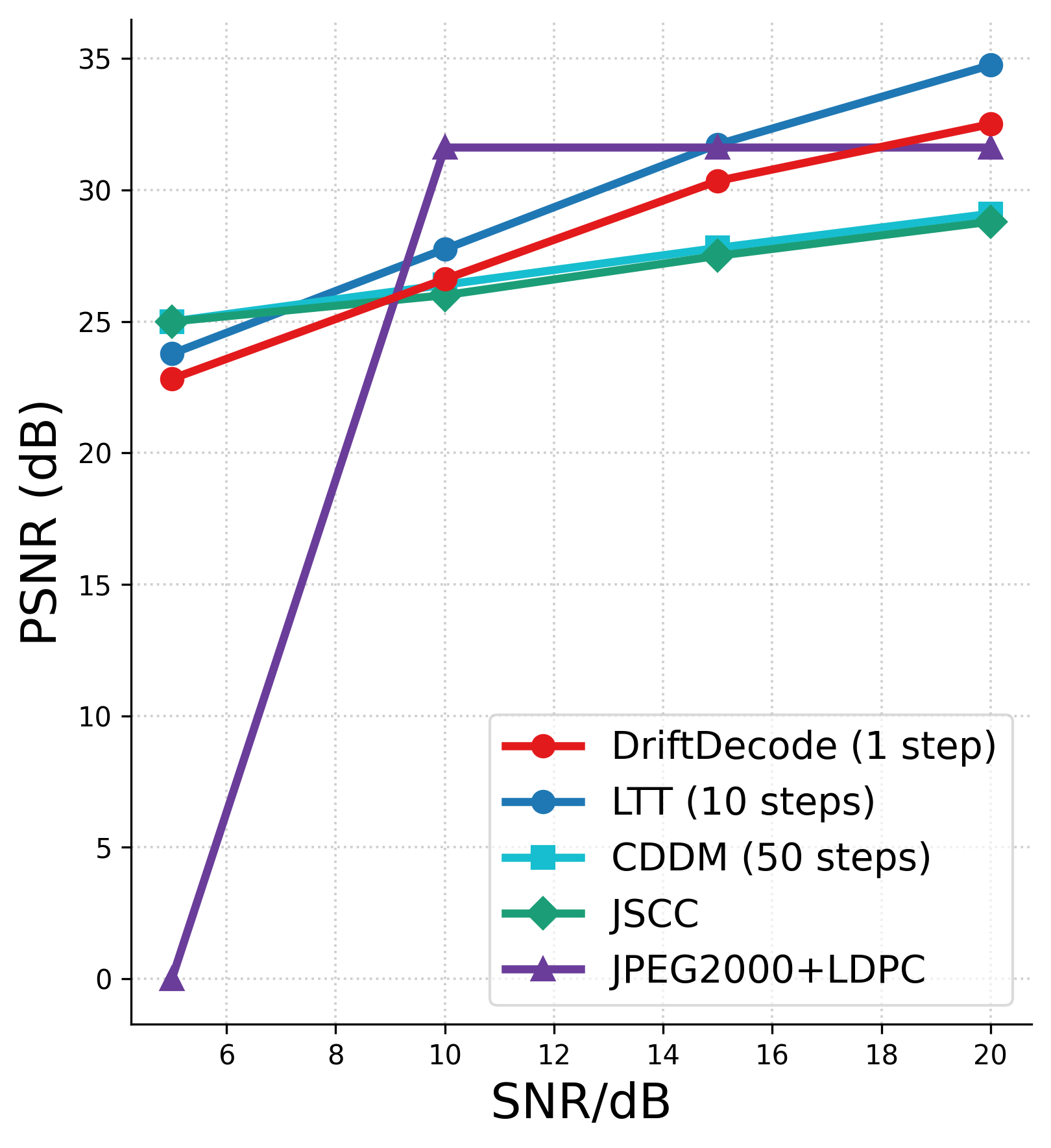}
        \caption{Rayleigh - PSNR}
        \label{fig:baseline_rayleigh_psnr}
    \end{subfigure}
    \hfill
    \begin{subfigure}[t]{0.24\textwidth}
        \centering
        \includegraphics[width=\textwidth]{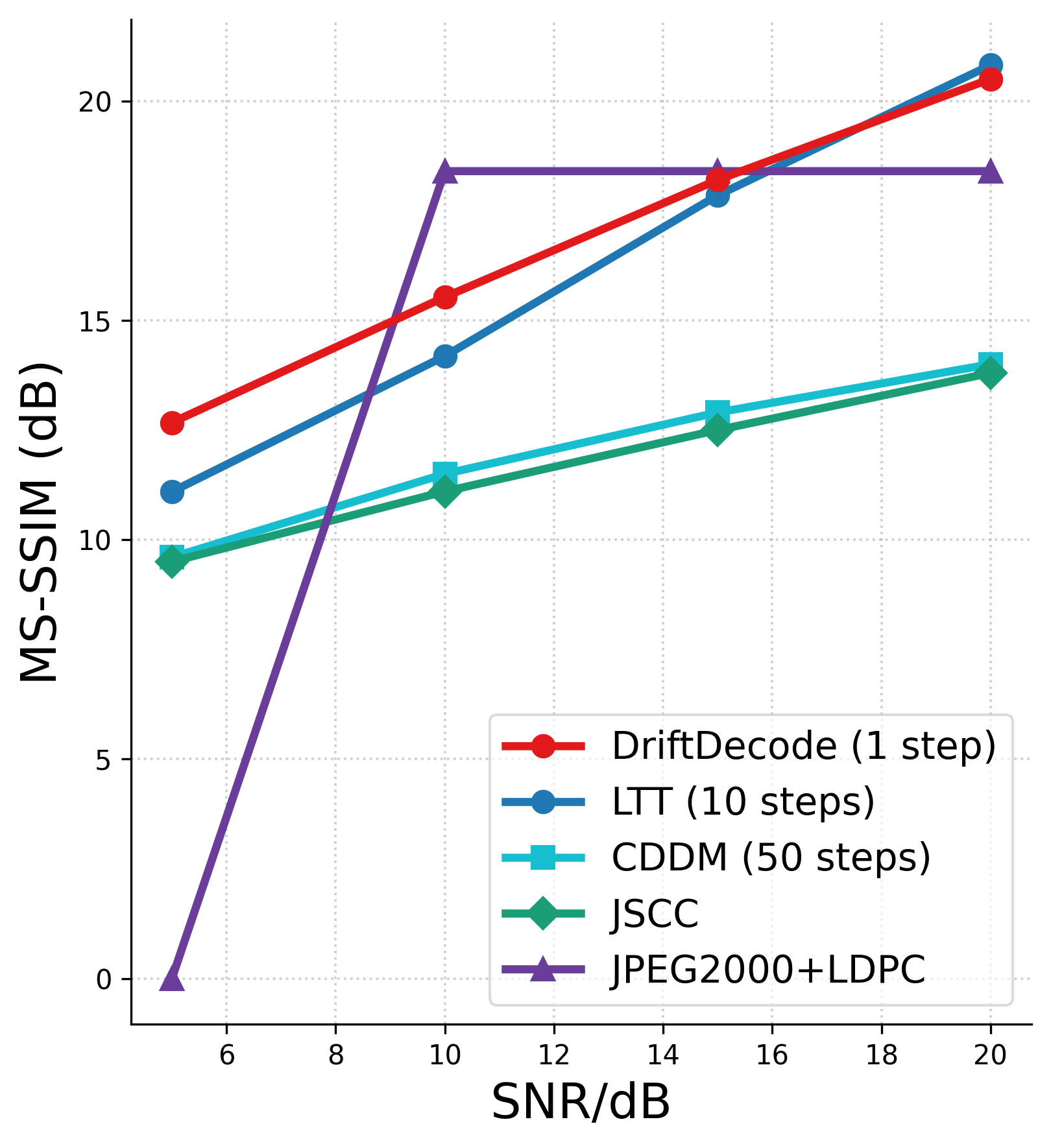}
        \caption{Rayleigh - MS-SSIM}
        \label{fig:baseline_rayleigh_msssim}
    \end{subfigure}
    \caption{Performance compared with baseline models in AWGN and Rayleigh channels on the DIV2K dataset.}
    \label{fig:four_images}
\end{figure*}
\subsection{Experimental Setups}
\subsubsection{Dataset and Training Protocol}
Experiments are conducted on two datasets: (1) DIV2K~\cite{agustsson2017div2k}, a high-resolution natural image dataset with 800 training images and 100 validation images, and (2) MNIST, a grayscale handwritten digit dataset with 60,000 training images and 10,000 test images at $32\times32$ resolution. For DIV2K, training and evaluation are performed on $256\times256$ image crops, and random cropping and horizontal flipping are applied for data augmentation. Models are trained for 100 epochs with batch size 8 (DIV2K) or 128 (MNIST) using AdamW ($\beta_1{=}0.9$, $\beta_2{=}0.999$) and weight decay 0.01. The learning rate is set to $10^{-3}$ with warmup, gradient norm is clipped to 2.0, and an exponential moving average (EMA) with decay 0.9995 is maintained. To enhance robustness across channel conditions, SNR is uniformly sampled from $[0,20]$~dB during training. Unless otherwise specified, the drift loss uses VGG-19 layers `conv2\_2', `conv3\_4', and `conv4\_4' with temperatures $\mathcal{T}=\{0.05, 0.2\}$. The drift loss is computed independently for each image using spatial locations within the same feature map. The loss weights are set to $\lambda_{\mathrm{mse}}=5.0$ and $\lambda_{\mathrm{drift}}=0.15$ unless otherwise specified. For fair comparison with LTT, the channel bandwidth ratio is fixed to $\mathrm{CBR}=1$ in all experiments.

\subsubsection{Baselines} Comparisons are made against four baselines: JPEG2000+LDPC as a separated source/channel coding pipeline, DeepJSCC~\cite{bourtsoulatze2019deep} as an end-to-end JSCC method, CDDM~\cite{wu2024cddm} as a diffusion-based decoder with approximately 50 sampling steps, and LTT~\cite{fu2026land} as a flow-matching decoder with 10 inference steps. For fair comparison among learned receivers, DriftDecode, DeepJSCC, CDDM, and LTT are compared under a unified setting with the same training data, the same $\mathrm{CBR}=1$ setting, and the same evaluation protocol.

\subsubsection{Evaluation Metrics}
Performance is evaluated using peak signal-to-noise ratio (PSNR), multi-scale structural similarity (MS-SSIM) \cite{wang2003multiscale}, and learned perceptual image patch similarity (LPIPS). PSNR is reported in dB, MS-SSIM is reported in dB as $-10\log_{10}(1-\mathrm{MS\mbox{-}SSIM})$, and LPIPS is reported in its original scale. Higher PSNR/MS-SSIM and lower LPIPS indicate better performance. All baseline models and DriftDecode are evaluated under the same protocol.

\subsection{Performance Comparison with Baseline Models}
Fig.~\ref{fig:four_images} compares DriftDecode with JPEG2000+LDPC, DeepJSCC, and generative-model-based receivers under AWGN and Rayleigh channels. The separated JPEG2000+LDPC pipeline shows a clear cliff effect: decoding collapses at low SNR (e.g., 5~dB) and then quickly saturates once above its operating threshold, providing limited graceful degradation. In contrast, learning-based schemes improve smoothly with SNR, which is more desirable for time-varying wireless links.
Across SNRs, DriftDecode consistently improves over DeepJSCC in both PSNR and MS-SSIM. This trend indicates that drift regularization enhances detail recovery beyond pixel-domain MSE training. Compared with the diffusion-based CDDM, DriftDecode achieves higher fidelity while avoiding iterative sampling, showing that multi-step generative inference is unnecessary when the received signal already carries substantial source information. Relative to the recent flow-matching receiver LTT, DriftDecode reduces the number of decoder evaluations from 10 to 1 while maintaining favorable PSNR/MS-SSIM trends; Table~\ref{tab:latency} further shows a 4.8$\times$ latency reduction. The results in Fig.~\ref{fig:four_images} indicate that one-step detail recovery provides a favorable quality-latency tradeoff for wireless image reconstruction.

\subsection{Performance on MNIST Dataset}
\label{sec:mnist}

The proposed DriftDecode method is also evaluated on MNIST under both AWGN and Rayleigh fading channels. As shown in Table~\ref{tab:mnist}, adding the drift loss consistently improves reconstruction quality at all SNRs, with gains that are modest under AWGN but substantially larger under Rayleigh fading. This trend indicates that drift regularization is particularly beneficial when channel distortion is more severe. A plausible explanation is that Rayleigh fading introduces amplitude-varying distortion across channel realizations, producing more diverse detail degradation patterns that benefit from the position-aware supervision of the drift loss.
\begin{table}[t]
\centering
\caption{MNIST results under AWGN and Rayleigh fading channels. $\Delta$PSNR and $\Delta$LPIPS denote the gain of MSE+Drift over MSE-only.}
\label{tab:mnist}
\begin{tabular}{c cc cc cc}
\toprule
\textbf{SNR} & \multicolumn{2}{c}{\textbf{PSNR (dB)}$\uparrow$} & \multicolumn{2}{c}{\textbf{LPIPS}$\downarrow$} & \textbf{$\Delta$PSNR} & \textbf{$\Delta$LPIPS} \\
\cmidrule(lr){2-3} \cmidrule(lr){4-5}
(dB) & MSE & +Drift & MSE & +Drift & (dB) & (\%) \\
\midrule
\multicolumn{7}{c}{\textbf{AWGN}} \\
\midrule
0  & 24.52 & \textbf{24.67} & 0.0490 & \textbf{0.0470} & +0.15 & $-$4.1 \\
5  & 27.93 & \textbf{28.06} & 0.0300 & \textbf{0.0290} & +0.13 & $-$3.3 \\
10 & 31.23 & \textbf{31.44} & 0.0180 & \textbf{0.0170} & +0.21 & $-$5.6 \\
15 & 34.87 & \textbf{35.04} & 0.0090 & \textbf{0.0080} & +0.17 & $-$11.1 \\
\midrule
\multicolumn{7}{c}{\textbf{Rayleigh}} \\
\midrule
0  & 22.27 & \textbf{23.16} & 0.0980 & \textbf{0.0720} & +0.89 & $-$26.5 \\
5  & 25.49 & \textbf{26.49} & 0.0630 & \textbf{0.0430} & +1.00 & $-$31.7 \\
10 & 28.78 & \textbf{29.91} & 0.0390 & \textbf{0.0240} & +1.13 & $-$38.5 \\
15 & 32.30 & \textbf{33.40} & 0.0210 & \textbf{0.0140} & +1.10 & $-$33.3 \\
\bottomrule
\end{tabular}
\end{table}

\subsection{Ablation Study: Loss Function}
\label{sec:ablation}

Table~\ref{tab:ablation_loss} reports the impact of different training objectives under the AWGN channel at 15~dB SNR. The MSE baseline achieves strong pixel-domain fidelity but tends to oversmooth high-frequency textures, which is a well-known limitation of MSE-based training~\cite{blau2018perception}. Adding LPIPS improves perceptual quality but provides limited gains in PSNR and no gain in MS-SSIM. In contrast, incorporating the proposed drift regularization with $\lambda_{\mathrm{drift}}{=}0.15$ yields the best overall tradeoff across PSNR, MS-SSIM, and LPIPS. The improvement is attributed to the position-aware drift field, which provides more targeted supervision for detail recovery than a generic perceptual loss~\cite{zhang2018lpips}.

\begin{table}[t]
\centering
\scriptsize
\caption{Loss ablation on DIV2K over the AWGN channel at 15~dB SNR (higher value $\uparrow$ is better for PSNR/MS-SSIM (dB), lower value $\downarrow$ is better for LPIPS).}
\label{tab:ablation_loss}
\begin{tabular}{l c c c}
\toprule
\textbf{Loss} & \textbf{PSNR (dB)} $\uparrow$ & \textbf{MS-SSIM (dB)} $\uparrow$ & \textbf{LPIPS} $\downarrow$ \\
\midrule
MSE only & 31.62 & 18.21 & 0.1640 \\
+ LPIPS & 31.77 & 18.21 & 0.1520 \\
\textbf{+ Drift} & \textbf{31.98 ($\uparrow$0.36)} & \textbf{18.38 ($\uparrow$0.17)} & \textbf{0.1420 ($\downarrow$0.0220)} \\
\bottomrule
\end{tabular}
\end{table}

\subsection{Ablation on Drift Loss Weight }

Table~\ref{tab:ablation_lambda} studies the effect of the drift regularization weight at 15~dB SNR. Performance improves consistently as $\lambda_\text{drift}$ increases from 0 to 0.15, indicating that moderate drift supervision enhances texture recovery beyond pixel-domain training. The best results are achieved at $\lambda_\text{drift}=0.15$. Further increasing the weight to 0.20 or 0.50 slightly degrades performance, suggesting that overly strong drift regularization can over-constrain the reconstruction.

\begin{table}[t]
\centering
\caption{Drift loss weight ablation on DIV2K at 15~dB SNR.}
\label{tab:ablation_lambda}
\begin{tabular}{c c c c}
\toprule
$\lambda_\text{drift}$ & \textbf{PSNR (dB)}$\uparrow$ & \textbf{MS-SSIM (dB)}$\uparrow$ & \textbf{LPIPS}$\downarrow$ \\
\midrule
0.00 (baseline) & 31.62 & 18.21 & 0.1640 \\
0.05 & 31.72 & 18.28 & 0.1460 \\
0.10 & 31.92 & 18.34 & 0.1440 \\
\textbf{0.15} & \textbf{31.98} & \textbf{18.38} & \textbf{0.1420} \\
0.20 & 31.94 & 18.33 & 0.1430 \\
0.50 & 31.86 & 18.25 & 0.1450 \\
\bottomrule
\end{tabular}
\end{table}

\subsection{Inference Latency}
\label{sec:latency}

Table~\ref{tab:latency} compares decoder-side inference cost for $256{\times}256$ images. LTT and DriftDecode are timed on the same NVIDIA A40 GPU. DriftDecode achieves 30~ms per image with a single forward pass, requiring 159~GFLOPs. In contrast, LTT relies on a 10-step iterative transport process and incurs 143~ms latency with 622~GFLOPs, yielding a 4.8$\times$ latency reduction and a 3.9$\times$ GFLOPs saving for DriftDecode. The diffusion-based CDDM is substantially more expensive due to iterative sampling, and its latency is reported as an estimate rather than a direct measurement. These results confirm that one-step decoding significantly reduces decoder latency while maintaining competitive reconstruction quality.

\begin{table}[t]
\centering
\caption{Decoder inference comparison for $256{\times}256$ images. LTT and DriftDecode were measured on an NVIDIA A40 GPU. CDDM values are estimated from its DDPM-style U-Net~\cite{wu2024cddm}.}
\label{tab:latency}
\begin{tabular}{l c c c}
\toprule
\textbf{Method} & \textbf{Steps} & \textbf{GFLOPs} & \textbf{Latency} \\
\midrule
CDDM~\cite{wu2024cddm} & $\sim$50 & $\sim$3100 & $\sim$714 ms \\
LTT~\cite{fu2026land}  & 10 & 622 & 143 ms \\
\textbf{DriftDecode (Ours)} & \textbf{1} & \textbf{159} & \textbf{30 ms} \\
\bottomrule
\end{tabular}
\end{table}

\section{Conclusion}
\label{sec:conclusion}

This paper proposed DriftDecode, an SNR-conditioned one-step decoder for wireless image reconstruction. The proposed framework treated wireless decoding as a detail recovery problem and reformulated the drifting-field mechanism as an instance-level texture loss in perceptual feature space to improve channel-impaired local detail restoration. Experimental results on DIV2K and MNIST under AWGN and Rayleigh fading channels showed that DriftDecode consistently improved upon MSE-only training while substantially reducing receiver-side decoding latency relative to a multi-step flow-matching baseline. These results support recovery-oriented one-step decoding as an effective design direction for low-latency wireless image transmission. Future work will consider bandwidth-limited regimes and robustness under imperfect CSI.


\bibliographystyle{IEEEtran}
\bibliography{refs}

\end{document}